\documentclass[useAMS,usenatbib]{mn2e}
\usepackage{epsfig}

\newcommand{\co}{\mbox{$^{12}$CO}}
\newcommand{\coa}{\mbox{$^{13}$CO~}}

\newcommand{\kms}{\mbox{km s$^{-1}$}}

\newcommand{\cc}{\mbox{cm$^{-3}$}}

\begin{document}

\title[Trans-Alfv\'{e}nic Motions in the  Taurus Molecular Cloud]{Trans-Alfv\'{e}nic Motions in the  Taurus Molecular Cloud}
\author[M.H. Heyer and C.M. Brunt]
{M. H. Heyer$^1$
and C. M. Brunt$^2$\\
$^{1}$Department of Astronomy, University of Massachusetts, Amherst, MA 01003, USA\\
$^{2}$School of Physics, University of Exeter, Exeter, EX4 4QL, UK}
\date{\today}



\maketitle

\label{firstpage}

\begin{abstract}
Magnetically aligned velocity anisotropy over varying physical conditions and environments within the Taurus Molecular Cloud 
is evaluated from analysis of  wide field spectroscopic imaging of \co\ and \coa\ J=1-0 emission.  Such anisotropy is a result of MHD turbulence in the strong magnetic field regime and provides an indirect measure of the role of magnetic fields upon the gas.   Velocity anisotropy aligned with the local, projected mean magnetic field direction  is limited to fields with low surface brightness \co\ emission corresponding to regions of low visual extinction and presumably, low gas volume density.  The more optically thin \coa\ J=1-0 emission shows little 
evidence for velocity anisotropy.  We compare our results with computational simulations with varying 
degrees of magnetic field strength and Alfv\'{e}nic Mach numbers.  In the diffuse, molecular  envelope of the cloud,  a strong magnetic 
field and sub-Alfv\'{e}nic turbulent motions are inferred.  Super-Alfv\'{e}nic motions are present within the 
high column density filaments of the Taurus cloud.   From this trans-Alfv\'{e}nic flow, we constrain the scaling exponent, $\kappa$, of the magnetic field 
density relation ($B \sim n^\kappa)$   to be near zero as expected for ambipolar diffusion or material loading of magnetic flux tubes. 
\end{abstract}

\begin{keywords}
MHD -- turbulence -- ISM: magnetic fields -- ISM: molecules -- ISM: kinematics and dynamics
\end{keywords}

\section{Introduction}
The origin of the supersonic motions observed within molecular clouds is a fundamental but poorly understood aspect of interstellar matter and star formation.  While such motions are attributed to turbulent magneto-hydrodynamic (MHD) flows within the cloud, it is currently uncertain whether velocities of such flows are less than (sub-Alfv\'{e}nic) or greater than (super-Alfv\'{e}nic) the Alfv\'{e}n velocity, $v_A$, in the cloud. This distinction is important in the context of describing
the formation of molecular clouds and the interstellar processes that regulate the production of stars.  Turbulent fragmentation considers supersonic, super-Alfv\'{e}nic turbulence that produces strong shocks at the interface of converging flows.  The resultant shock-induced density perturbations are seeds to gravitationally unstable cores from which individual stars and clusters condense (Padoan \& Nordlund 2002;  Krumholz \& McKee 2005).  
For sub-Alfv\'{e}nic clouds, the observed motions are  attributed to a spectrum of Alfv\'{e}n waves (Arons \& Max 1975; Kudoh \& Basu 2003; Mouschovias, Ciolek, \& Morton 2011).  
The allowed range of Alfv\'{e}n wavelengths are bounded on the large scale by the size of the cloud and on the small scale by the ion-neutral mean free path (McIvor 1977; Carlberg \& Pudritz 1990).   
Observationally, the Alfv\'{e}nic character of molecular clouds is ill-defined. Recent compilations of Zeeman measurements of the HI 21cm line in absorption against continuum sources by Heiles \& Troland  (2005) and OH emission toward
molecular cloud cores by Troland \& Crutcher (2008) found $M_A = \sigma_v/v_A \sim$1-2.  However, optical and infrared 
polarization studies typically measure a uniform field  geometry in cloud envelopes (Moneti et al. 1984; Heyer et al 1987; Goodman et al. 1990), which indicates a field resistant to 
entanglement by strong turbulent eddies  (Chandrasekhar \& Fermi 1953) and sub-Alfv\'{e}nic turbulence.  

An indirect measure of the influence of the interstellar magnetic field upon the cold, neutral gas component is 
the degree of velocity anisotropy aligned with the magnetic field that can result from MHD turbulence. 
 In this context, velocity anisotropy refers to 
the difference in directional power spectra or structure functions of the velocity field along two orthogonal directions (Goldreich \& Sridhar 1995).
Such magnetically aligned velocity anisotropy can be induced by the interactions between three or more Alfv\'{e}n wave packets on localized scales corresponding to 
fluctuations of the mean field component, ${\delta}B$,   as proposed 
by Goldreich \& Sridhar (1995).   On larger scales,  a strong mean field threading an interstellar cloud  
may also regulate the flow of material  (Cho \& Lazarian 2003).  Computational simulations demonstrate 
that magnetically aligned velocity anisotropy occurs only when the field is strong ($c^2/v_A^2$=0.01) and the Aflvenic 
Mach number is less than unity, ($M_A < 1$)   (Cho \& Lazarian 2003;  Vestuto, Ostriker, \& Stone 2003).  More recently,  Esquivel \& Lazarian (2011) used velocity centroids derived from numerical MHD simulations to further confirm a relationship between velocity anisotropy and gas magnetization (B, $M_A$).  They also inferred a dependence of the measured anisotropy with the sonic Mach number for highly magnetized gas. 

Using the simulations of Ostriker, Stone, \& Gammie (2001), Heyer etal (2008)  developed and calibrated an
analysis technique (see $\S2$) that derives pseudo velocity structure functions along two orthogonal axes from 
imaging data of spectral line emission.  Transforming the set of velocity and density fields generated by the MHD simulations to model observations of \co\ and \coa\ J=1-0 emission and applying their analysis to the model data set, they verified that observational signatures of magnetically aligned velocity anisotropy are limited to the regime at which the field is strong and the motions are sub-Alfv\'{e}nic.    
Heyer etal (2008)  then applied this method to a subfield of imaging observations of \co\ and \coa\  J=1-0 line emission from the Taurus molecular cloud (Goldsmith etal 2008; Narayanan etal 2008).  
They identified velocity anisotropy that is aligned with the local 
magnetic field direction.  Assuming a mean density and gas temperature and comparing the measured 
degree of anisotropy with values collected from model observations derived from the computational simulations, they derived a magnetic field strength of 14 $\mu$G 
and a subcritical ($\Sigma_{gas}/\Sigma_{crit} < 1$) cloud envelope, where $\Sigma_{gas}$ is the central gas surface 
density and $\Sigma_{crit} = \langle B \rangle/(63G)^{1/2}$ is the surface density that parametrizes magnetic support of the gas against self-gravity with mean field $\langle B \rangle$
(Mouschovias \& Spitzer 1976).  

The sub-field within the Taurus cloud analyzed by Heyer etal (2008) is not an unbiased target.  The 
field was selected for this analysis owing to the presence of striations of \co\ emission that are parallel with 
the local magnetic field direction as derived from optical polarization of background stars that assumes 
the alignment of elongated grains with the field.  Correspondingly, there was a reasonable expectation that the magnetic field 
plays some role in the gas distribution and kinematics in this selected region of the cloud.   Given this biased and limited area, the more general
influence of the static magnetic field component or magnetic 
wave phenomena on the gas throughout the Taurus cloud has not yet been fully  assessed.   

To more fully evaluate the role of the interstellar magnetic field and its variation throughout the Taurus molecular cloud complex, we have extended the study of Heyer etal (2008) by analyzing the presence of velocity anisotropy within 180 partitioned subfields of the Taurus Survey map.   Specifically, we are interested in the conditions over which 
the magnetically aligned velocity anisotropy varies within the cloud.  Such variations can occur as the cloud ionization
changes from the envelope to the more deeply-embedded filaments and core regions or via a transition from sub-Alfv\'{e}nic to super-Alfv\'{e}nic turbulence.   In $\S2$, we briefly summarize the molecular line data and the 
basic analysis, both of which are respectively described in more detail in earlier publications (Narayanan etal 2008; 
Heyer etal 2008).  In $\S3$, we apply this analysis to the partitioned Taurus field for both \co\ and \coa\ data sets.
The results are placed into a broader physical context in $\S4$.

\section{The Data and Analysis Method}
All molecular line data used  in this study are part of the FCRAO Survey of the Taurus Molecular Cloud described by 
Narayanan etal (2008) and  Goldsmith etal (2008).    The survey imaged \co\ and \coa\ J=1-0 emission from
96 deg$^2$ of the Taurus cloud with the FCRAO 14 meter telescope.  The frontend consisted of the  
32 element focal plane array receiver, SEQUOIA that fed a set of autocorrelation backend spectrometers
configured with a spectral resolution of 25 kHz or $\sim$0.06 \kms\ for both observed transitions.
The Taurus data used in this study have been further processed to account for contributions and losses from the antenna error beam (Pineda et al. 2010).  The result of this deconvolution is a calibration of antenna temperatures that, when smoothed to the same angular resolution, are congruent with values from the CfA 1 meter telescope for which such error beam contributions are minimal.   In addition, the \co\ and \coa\ spectra have been 
resampled onto the same spectral grid with 0.1 \kms\ spacing.  

The calculation of pseudo-structure functions along two perpendicular  axes is described by Heyer etal (2008). 
A  set of eigenvectors and eigenprojections are derived along a sequence of 
position-velocity slices of a spectroscopic data cube using Principal Component Analysis along each cardinal 
direction of the spectroscopic data cube.  Such a decomposition 
measures the spatial variation of line profiles along the axis as these differences generate variance.  From the 
eigenvectors and eigenprojections that account for signal variance above values expected from thermal noise, 
one can respectively derive the velocity difference and scale over which the difference occurs in order to build a
pseudo-structure function.  We refer to these measures as pseudo-structure functions because our analysis does 
not directly sample velocity differences from the true 3 dimensional velocity field but rather,  extracts velocity displacements from the differences in the  spectral line profile shapes that are an integration of density and projected velocities along a line of sight through the cloud.    The pseudo structure functions are derived for the two cardinal directions of the 
data cube.  The velocity anisotropy is measured from the differences between the power law parameters that characterize the pseudo-structure functions on each axis.    To find the angle of maximum velocity anisotropy, $\theta_{MA}$, the data cube is rotated through a sequence of angles, 
from 0 to 180 degrees spaced by 10 degrees.  For each angle, the measure of velocity anisotropy, $\Psi(\theta)$,  is calculated.   This modulation allows one to fit for the phase, $\theta_{MA}$ and amplitude, $\Psi_\circ$, for which the anisotropy is a maximum.   The value of 
$\theta_{MA}$ can be compared to the local magnetic field direction, $\theta_B$, as determined by optical and infrared 
polarization of background stars.   

\section{Results} 
To assess the degree of velocity anisotropy throughout the Taurus Molecular Cloud, we have partitioned 
the field surveyed by Narayanan etal (2008) into 180 blocks, each with 256$\times$256 pixels.  This size corresponds to 1.42$\times$1.42 deg$^2$ or 3.5$\times$3.5 pc$^2$ at a distance of 140 pc.  The blocks are  spaced 
by 128 pixels in both the R.A. and Declination directions.   The selected size of each partition is a compromise
between having a field sufficiently large to derive useful statistics along any axis while limiting the area over which significant variations of the magnetic field geometry may occur, which can dilute any signature to MHD velocity anisotropy (Cho \& Lazarian 2003).   

\begin{figure*}
\begin{center}
\leavevmode
\epsfxsize=15cm\epsfbox{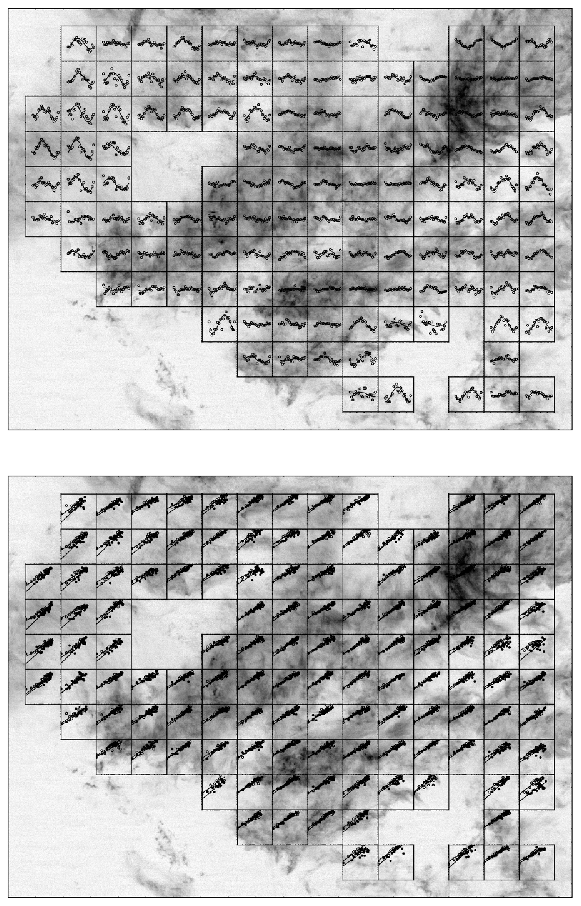}
\caption{(top) The variation of the anisotropy index, $\Psi(\theta)$ with rotation angle of the data cube for each sub-field with sufficient signal to carry out the analysis using the \co\ J=1-0 data.
The solid line shows the fit of the points to the function $\Psi=\Psi_\circ cos[2(\theta-\theta_{MA})]$.
(bottom)  The pseudo velocity structure 
functions along the cardinal axes (solid circles: x axis; open circles: y axis) evaluated at the angle of maximum anisotropy, $\theta_{MA}$. 
}
\label{fig1}
\end{center}
\end{figure*}

\begin{figure*}
\begin{center}
\leavevmode
\epsfxsize=15cm\epsfbox{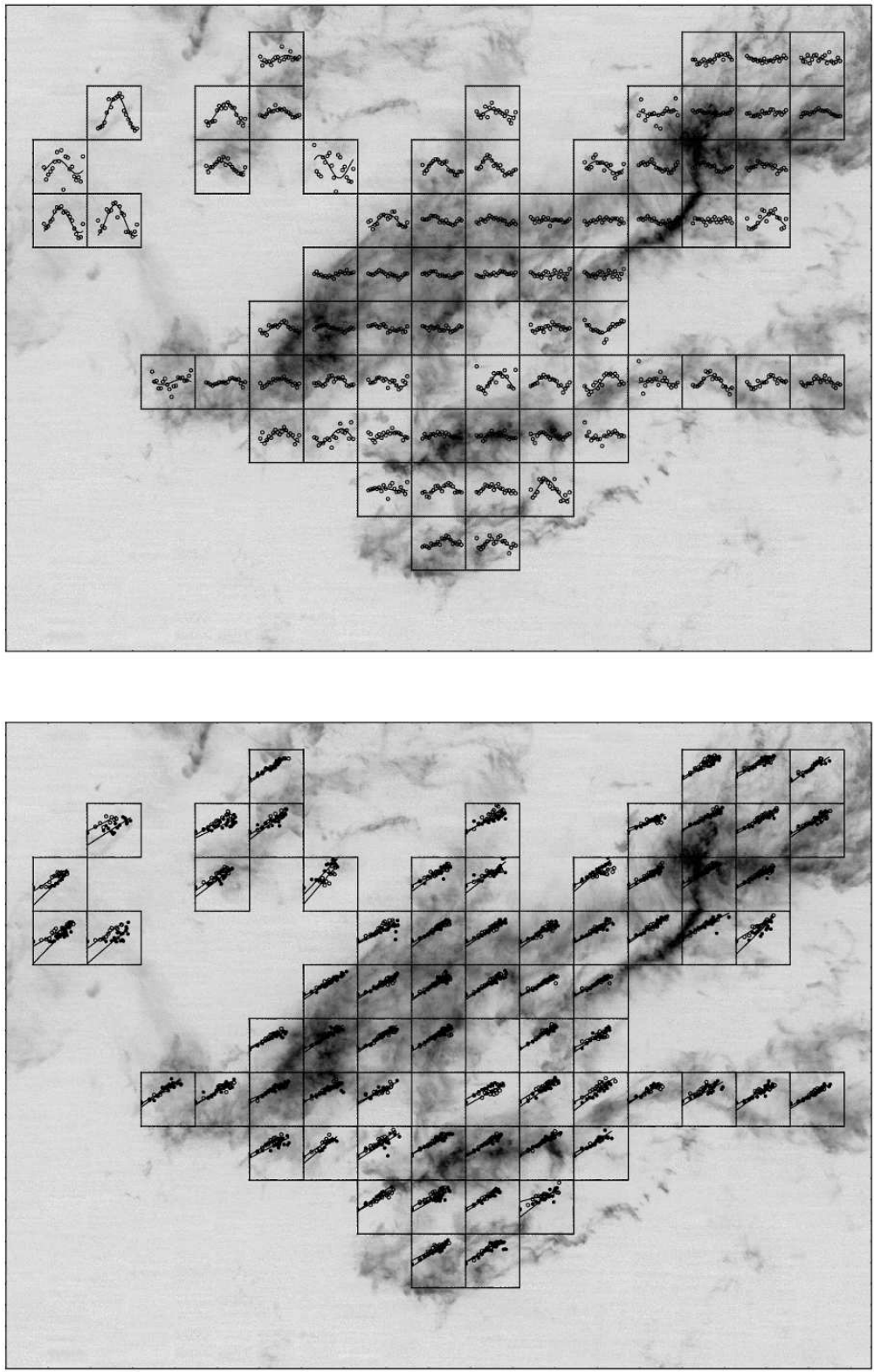}
\caption{Same as Figure~\ref{fig1} with \coa\ J=1-0 data. 
}
\label{fig2}

\end{center}
\end{figure*}

For each partition,  the \co\ and \coa\ emission are analyzed for velocity anisotropy following the procedure described by Heyer etal (2008).   In this study, we exclusively use the normalized differences between the 
power law amplitudes as a measure of velocity anisotropy as these parameters are more reliably determined in the fits to the pseudo-structure functions than the power 
law indices given the limited spatial dynamic range afforded by the position-velocity images.   
The results of this analysis are shown  in Figure~\ref{fig1} and Figure~\ref{fig2}  that show the variation
of the anisotropy index, $\Psi(\theta)$ with $\theta$ and the perpendicular pseudo-structure functions evaluated at the 
angle of maximum anisotropy, $\theta=\theta_{MA}$, for \co\ and \coa\ respectively. 
Strong velocity anisotropy is identified as 
a strong modulation of $\Psi(\theta)$ with $\theta$ and displaced pseudo-structure 
functions along the two perpendicular axes evaluated at $\theta=\theta_{MA}$.   
Partitions for which the calculations are not well defined owing to a limited spatial dynamic range with line emission are not shown.  Such conditions are realized in partitions where there is little or no detected 
emission or when the partition is positioned on the edge of an emission feature that truncates the 
available scales to analyze. Since the \co\ emission extends over larger scales than \coa\ emission owing to a 
larger abundance within regions of low extinction and radiative trapping, there are 
more partitions with valid measures of the velocity anisotropy.  

The strong velocity anisotropy identified by Heyer etal (2008) in the 
region of emission striations in the northeast sector of the image is confirmed within the individual partitions that comprise this area.  In comparison with the 
rest of the cloud, this area exhibits the largest anisotropy amplitudes.   Other notable regions with strong 
velocity anisotropy include partitions along the northern and western edges of the Taurus cloud.  

\begin{figure*}
\begin{center}
\leavevmode
\epsfxsize=15cm\epsfbox{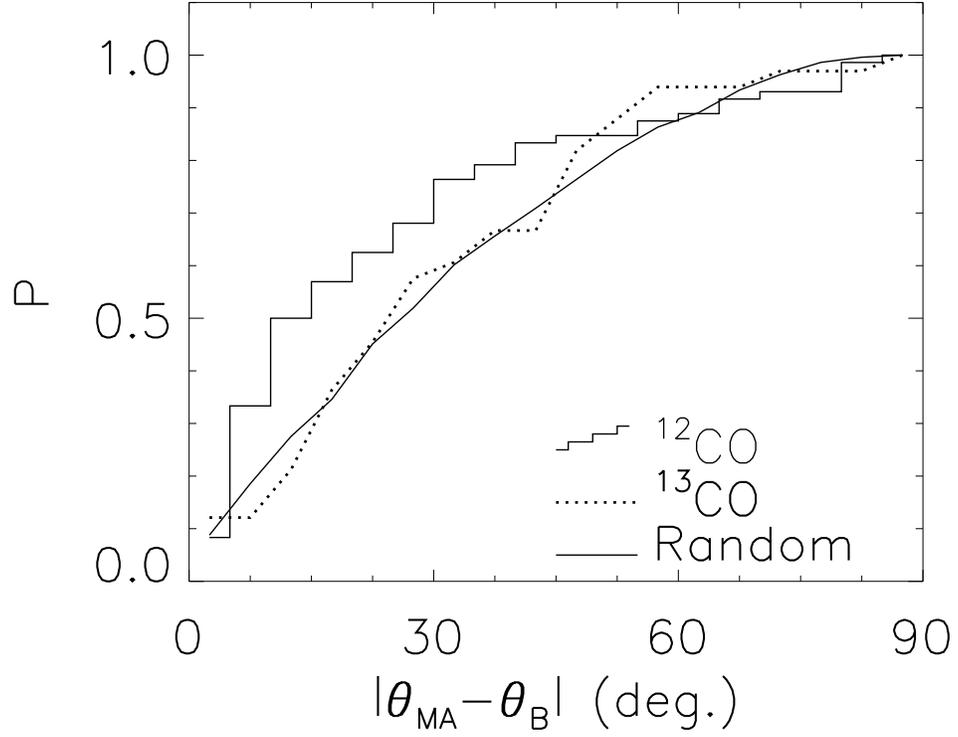}
\caption{The cumulative distribution of the absolute value of the difference between 
the angle of maximum anisotropy and the local magnetic field direction for 
\co\ (histogram), \coa (dotted line), and a random distribution of angle differences (solid line).  
}
\label{fig3}
\end{center}
\end{figure*}

For both \co\ and \coa, velocity anisotropy is only detected in regions of low surface brightness emission.  To link this anisotropy to the interstellar magnetic field that threads the 
cloud, such anisotropy should be approximately aligned with the local field direction.  
To assess the degree of alignment, we have examined the local magnetic field direction in each partition using 
the set of infrared and optical polarization vectors (Moneti et al. 1984; Heyer et al. 1987; Goodman et al. 1990; Whittet et al 1992).  
 For each partition block, we calculate 
a mean polarization angle, $<\theta_B>$, derived by weighting each angle by the degree of polarization
\begin{equation}
 < \theta_B> = \Sigma \theta_i P_i / \Sigma P_i  
\end{equation}
where the sum is over all polarization measurements within the subfield and  $\theta_i$ and $P_i$ are the polarization angle and percent polarization for the $i^{th}$ measurement within the partition.  A minimum of two polarization measurements within a subfield is required to derive $<\theta_B>$.   
Figure~\ref{fig3} shows the cumulative distribution of the absolute value of the displacement between 
the angle of maximum anisotropy and the mean local magnetic field direction, $|\theta_{MA}-<\theta_B>|$,
for both \co\ and \coa\  data sets.  Also shown is the cumulative distribution expected for a uniformly random distribution of 
angle displacements.    The angular displacements derived from the \coa\ data appear  consistent with a sample of 
values drawn from a random distribution.   This is not surprising since there is little 
velocity anisotropy recovered by the \coa\ data so the corresponding angle of maximum anisotropy is not well defined.  
The Kolmogorov-Smirnov (K-S) Test provides a quantitative measure of the likelihood that two distributions 
are drawn from the same population.  The Kolmogorov-Smirnov D statistic of the  \coa\ angular displacement and a random distribution is 0.162 corresponding to a probability of 36\% that the two distributions are similar.  In contrast, the distribution of 
angular displacements derived from the \co\ data shows an enhancement at small values of 
$|\theta_{MA}-<\theta_B>|$.  Comparing these values  to a random distribution, the Kolmogorov-Smirnov D statistic is  
0.292, which implies a very low probability (P=3$\times$10$^{-5}$ ) that the \co\ derived displacements  are drawn from a random distribution.  

\begin{figure*}
\begin{center}
\leavevmode
\epsfxsize=15cm\epsfbox{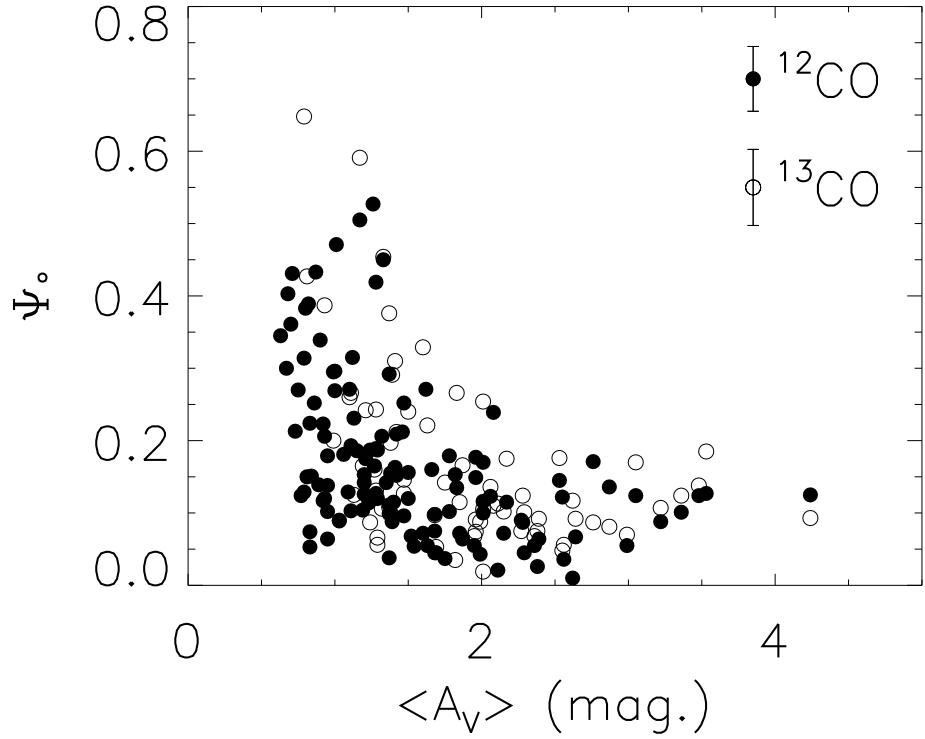}
\caption{(left) The variation of the amplitude of velocity anisotropy, $\Psi_\circ$
with the mean visual extinction over each subfield.  The solid and open circles 
are derived from \co\ and \coa\ respectively. 
Magnetically aligned velocity anisotropy is 
only evident within the low column density envelope regions of the Taurus cloud.
}
\label{fig4}
\end{center}
\end{figure*}

Figure~\ref{fig1} and Figure~\ref{fig2} illustrate that signatures to velocity anisotropy are exclusively limited to regions of low surface brightness \co\ and \coa\ line emission.   The low CO surface brightness emission corresponds to regions of low gas column density or visual extinction.  The faint signal for both \co\ and \coa\ from these lines of sight is due to subthermal excitation conditions corresponding to gas volume densities much lower than the critical density of CO (n $<$10$^3$ \cc).   To more 
directly establish this connection,  the variation of the amplitude of velocity anisotropy, $\Psi_\circ$, with visual extinction, $A_v$, in each partition is derived and shown in Figure~\ref{fig4}.   
The visual extinctions are derived from the 2MASS $A_v$ image
presented by Pineda etal (2010).
Typical error bars derived from the 
fitting errors of $\Psi(\theta)$ for \co\ and \coa\ are shown in the legend.  The extinctions displayed in Figure~\ref{fig4} are averaged over the 1.4$\times$1.4 deg$^2$ area of each partition.  At this resolution, the dynamic 
range of extinction values is much smaller than the original $A_v$ image of Pineda etal (2010).     The data 
show that significant velocity anisotropy is limited to regions with mean extinctions less than 2 magnitudes. 

\section{Discussion}
Velocity anisotropy aligned with the magnetic field is a basic 
feature of strong MHD turbulence in the interstellar medium
 (Kraichnan 1965; Montgomery \& Turner 1981; Higdon 1984; Goldreich
\& Sridhar 1995; Lithwick \& Goldreich 2002; Cho \& Vishniac 2000).  For most cases, the anisotropy results from the interaction between oppositely propagating Alfv\'{e}n wave packets that 
differentially distribute kinetic energy transverse and parallel to the local magnetic field. The earliest versions of these descriptions focused on 
density fluctuations in a highly ionized, magnetic medium inferred from scintillation
measures.  However,  in the regime of strong ion-neutral coupling, these processes should also be applicable to the cold, mostly neutral molecular ISM as traced by the molecular line emission used in this study subject to the limitation that the 
wavelength of any disturbance must be larger than the ion-neutral mean free path (Arons \& Max 1975).  

The degree of velocity anisotropy predicted by Goldreich \& Sridhar (1995) increases with decreasing scale as evaluated 
within the local frame of reference over which the magnetic field direction is uniform.  That is, turbulent eddies become more 
elongated as these interact with the local magnetic field.  Due to the effects of the Alfv\'{e}n waves, one would expect some degree of curvature and wandering of the field lines over various scales corresponding to the wavelengths of the disturbances.   Ideally, one would evaluate the velocity anisotropy within the local frame of reference.  Cho \& Vishniac
(2000) describe two methods to identify the local frame for any location within a 3 dimensional field as provided by 
computation simulations.   Evaluating eddy shapes within this frame of reference both Cho \& Vishniac (2000) and 
Cho, Lazarian, \& Vishniac (2002) confirm the GS95  scaling between parallel and perpendicular wave vectors. 
However, such methods are not applicable to real observations that are limited  to projected distributions of the magnetic field geometry on the plane of the sky.  While we have attempted to account for projected, large scale curvature of the 
magnetic field by deriving a local, mean magnetic field direction in each partition, these measures are not sensitive to deflections of the field geometry along the line of sight that must be present owing to the transverse perturbations of the Alfv\'{e}n waves themselves.   Nor do these account for any 
projected substructure within each 1.4$\times$1.4 deg$^2$ area.    Furthermore, since our method is 
sensitive to the field component projected onto the plane of the sky, any inclination of the magnetic field would further 
reduce the degree of measured anisotropy.   Yet, despite these geometric limitations, velocity anisotropy aligned along 
these projected, mean magnetic field directions is measured.  By not accounting 
for the inclination of the mean magnetic field relative to the plane of the sky and the  perturbations of the field 
direction along the line of sight, our measures of anisotropy are necessarily lower limits to the true degree of anisotropy.  
It is not clear whether these measures of anisotropy arise from the effects described by GS95 or are due to some other processes
that differentially regulate the cascade of energy with respect to the mean field direction within the partition area.  

The comparison of our method to measure velocity anisotropy with numerical simulations demonstrates the limited regime over which such magnetic asymmetry is present (Heyer etal 2008).  This regime is characterized by 
strong magnetic fields ((c/v$_A)^2 <<$ 1) and sub-Alfv\'{e}nic motions,  $\sigma_v/v_A < 1$.
Both conditions lead to  magnetic field lines that are resistant to being tangled by vortical motions such that the 
Alfv\'{e}n wave induced perturbations to the magnetic field, $\delta$B, are small with respect to the mean field, B$_\circ$. 
Based on this previous calibration effort, we infer that such strong magnetic conditions prevail in the regions in the Taurus 
cloud where clear evidence for magnetically aligned velocity anisotropy is identified.    These regions correspond to the molecular 
envelope of the cloud that is traced by the low surface brightness \co\ emission.    This component also shows 
evidence for magnetically aligned striations with patterns that are consistent with wave phenomena within the cloud.  

Differential behavior of the measured anisotropy is identified between \co\ and \coa.  The 
lower opacity \coa\ J=1-0 line effectively probes the column density range 4 $<$ A$_v$ $<$ 10 magnitudes
(Goldsmith etal 2008; Pineda etal 2010).
Owing to increased optical depth and chemical depletion of \co\ and \coa\  molecules onto grains within the 
cold, high density core regions, neither molecule reliably probes this high volume and column 
density regime of the Taurus cloud.  
For an optically thin line, low surface brightness emission generally corresponds to lower column density relative to that probed by high surface brightness signal.  For the optically thick 
\co\ line, the faint signal within this low surface brightness component is due to subthermal excitation conditions corresponding to 
low gas volume densities.   Given the excitation requirements and the limits imposed by opacity and chemical depletion, the range of volume densities responsible for 
the observed \co\ and \coa\ J=1-0 emissions is $10^2 < n < 7500$ \cc.   

We propose that the observed differential measurements of velocity anisotropy between the low 
and high surface brightness regions and \co\ and \coa\ emission reflect a 
transition of the MHD turbulent flow.  In the low 
volume density, low column density regime, the field is strongly coupled to the neutral 
material and the motions are sub-Alfv\'{e}nic.  In the high column density regime
of the cloud,  the gas motions are super-Alfv\'{e}nic.   The Alfv\'{e}nic Mach number in the Taurus envelope, $M_{A,0} $, is not 
directly established from our data.   Following Heyer etal (2008), we estimate $M_{A,0} \sim 0.5$ based on the comparison of the 
degree of anisotropy with the models.  Therefore, the Alfv\'{e}nic Mach number must increase by a factor greater than 2 between the 
envelope and dense filaments to account for the differential behavior of velocity anisotropy. 

The transition from sub-Alfv\'{e}nic motions in the cloud envelope to super-Alfv\'{e}nic velocities in 
the high column density interior of the cloud offers insight to the evolution of material and 
the role of the magnetic field.  
The 
Alfv\'{e}n velocity is expressed as $v_A=\langle B \rangle/\sqrt{4{\pi\mu}m_H n}$ where $\langle B \rangle$ is the mean magnetic field strength and n is the gas volume density 
within the volume $l^3$.   Owing to increasing density and the amplification of the magnetic field, $v_A$ is expected to 
vary between the envelope and the dense, filaments.  

A simple, heuristic scaling relationship is constructed  that describes the 
variation of the Alfv\'{e}nic Mach number, $M_A$ with the observed  physical properties of the cloud.  
The 
gas density, $n$ can be expressed  $n=\Sigma/l$, where $\Sigma$ is the gas mass surface density 
and $\l$ is the size scale of the volume, $l^3$. 
Both $\Sigma$ and $l$ are more accessible with 
observational methods than $n$.  The magnetic field, $\langle B \rangle$, scales with density as 
$\langle B \rangle=B_\circ(n/n_\circ)^\kappa$.   
 Combining these relationships, the Alfv\'{e}n velocity scales 
as 
\begin{equation}
v_A \sim \langle B \rangle/n^{1/2} = v_{A,\circ} (\Sigma/\Sigma_\circ)^{\kappa-1/2} (l_\circ/l)^{\kappa-1/2} 
\end{equation}
where $v_{A,\circ}$ is the Alfv\'{e}n velocity with conditions $\Sigma_\circ$ at size $l_\circ$.  
Any geometrical factors related to re-casting the volume density to $\Sigma/l$ are folded into $v_{A,\circ}$.     
The velocity dispersion
characterizing clouds motions within a volume, $l^3$ is described by the velocity structure function,
\begin{equation}
\sigma_v = \sigma_{v,0} (\Sigma/\Sigma_\circ)^{1/2} (l/l_\circ)^{\gamma}
\end{equation}
We include the dependence of the velocity structure function on the gas surface density {\em within a GMC}.
Heyer et al. (2009) identified a link between the gas surface density of a GMC and 
$\sigma_{v,L}/L^{1/2}$, where $\sigma_{v,L}$ is the velocity dispersion measured over the cloud size, $L$.   
If the power law index of the first order structure function, $\gamma$ = 1/2, as determined by Heyer \& Brunt (2004), 
then the quantity 
$\sigma_{v,L}/L^{1/2}$ provides the normalization of the velocity structure function of the GMC.   Therefore, we 
necessarily include the factor, $(\Sigma/\Sigma_\circ)^{1/2}$, in the expression for $\sigma_v$.   For the remainder 
of this paper, we will assume that $\gamma=1/2$. 
The Alfv\'{e}nic Mach number, $M_A$ then scales as
\begin{equation}
M_A = \sigma_v/v_A = \sigma_{v,0} (\Sigma/\Sigma_\circ)^{1/2} (\l/\l_\circ)^{1/2}  / v_A 
\end{equation}
\begin{equation}
M_A  = M_{A,0} (\Sigma/\Sigma_\circ)^{1-\kappa} (l/l_\circ)^\kappa 
\end{equation}
where $M_{A,0}$ is the Alfv\'{e}nic Mach number at conditions with surface density $\Sigma_\circ$ and size scale, $l_\circ$.  

\begin{figure*}
\begin{center}
\leavevmode
\epsfxsize=15cm\epsfbox{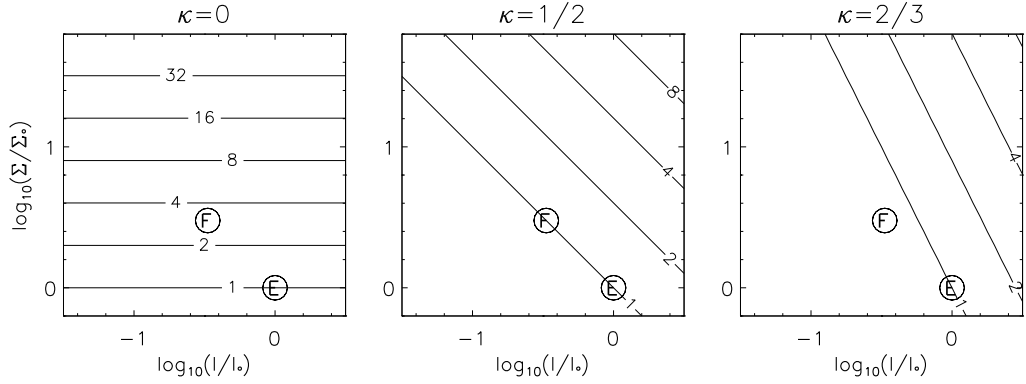}
\caption{Contours of $M_A/M_{A,\circ}$ values as a function of size scale, $l$ and surface density, $\Sigma$ normalized to 
initial values $l_\circ$ and $\Sigma_\circ$ for  $\kappa$ = 0 (left panel), 1/2 (middle panel), and 2/3 (right panel). 
The circumscribed symbols approximate the relative size scale and surface density  conditions 
between the low column density envelope (E) and the dense filaments (F) of the Taurus cloud.
A transition of sub-Alfv\'{e}nic 
to super-Alfv\'{e}nic motions ($M_A/M_{A,\circ} > 2$) between the cloud envelope and dense filaments is only possible for $\kappa=0$.
}
\label{fig5}
\end{center}
\end{figure*}

In Figure~\ref{fig5}, we show the variation of $M_A/M_{A,0}$ with size scale and surface density for three physical 
cases $\kappa = 0,1/2, 2/3$. 
The value of $\kappa=2/3$ corresponds to a spherically
contracting cloud with strict flux freezing conditions;  $\kappa=1/2$ is expected from an isothermal, gravitationally contracting fragment or core with frozen flux and  mass to flux ratio greater than the critical value, $(M/\Phi)_{cr} = 1/(63G)^{1/2}$ (Mouschovias \& Spitzer 1976).  
For material falling down magnetic field lines as in the case of ambipolar diffusion over the limited density range probed by  \co\ and \coa\ J=1-0 emission  (Fiedler \& Mouschovias 1993) or shocks aligned 
with the mean field field direction, $\kappa \sim 0$ (Hennebelle et al. 2008; Heitsch, Stone, \& Hartmann 2009). 
It is noteworthy that for $\kappa=0$,  $M_A/M_{A,0} $  has no dependence on size scale (horizontal 
lines) so that $M_A/M_{A,0} $ increases with {\em any} increase in surface density as expected 
between the cloud envelope and dense cores or filaments.  For the other cases
($\kappa=1/2, 2/3)$,  a much higher contrast of mass surface density ($\Sigma/\Sigma_\circ$) over a limited 
change in size scale  is required to significantly increase $M_A/M_{A,0} $.   The amplification of the 
field with increasing density constrains $v_A/v_{A,\circ} \ge 1$ while $\sigma_v/\sigma_{v,\circ}$ decreases due to the 
energy cascade of turbulence. 

A  visual inspection of the Taurus maps provides a coarse estimate of the conditions between the cloud envelope and 
the dense filaments: $l/l_\circ \le 1/3$ and 
$\Sigma/\Sigma_\circ \ge 3$. 
For these conditions, the Alfv\'{e}nic Mach number would remain constant between the envelope and dense filament regimes 
for $\kappa=1/2$ and would even decrease for $\kappa=2/3$. 
 These observational constraints exclude magnetic conditions for 
which $\kappa \ge 1/2$.   In contrast, the accumulation of material along magnetic field lines
without significant magnetic field amplification ($\kappa \sim 0$) can readily achieve the necessary 
factor, $M_A/M_{A,0} \ge 2$, within the observed conditions of the Taurus cloud. 

Near zero values of $\kappa$ over the density ranges corresponding to our \co\ and \coa\ observations 
are predicted for molecular clouds whose evolution is regulated by ambipolar diffusion or the large scale 
shocks perpendicular to the field that forces material along the mean field direction.   The latter 
scenario could occur as two clouds moving in opposite directions along the magnetic field  collide
(Heitsch, Stone, \& Hartmann 2009).   While this description is plausible, the requirement that the relative motions be 
strictly along
the magnetic field would make this an improbable event.  Even a small, angular offset between the ordered magnetic field 
direction and that of the converging flow would readily result in a field component parallel to the compressed layer that would be 
further amplified by the local dynamic pressure of the shock.  
 Alternatively,
ambipolar diffusion is an inevitable process in the regime of low ionization and an initially 
sub-critical cloud envelope.  A recent polarization study by Chapman et al. (2011) derived 
sub-critical conditions for several of the major subregions of the Taurus cloud over scales of 
$\sim$ 2 pc.   We propose that the observed velocity and column density 
structure of the Taurus cloud is a direct consequence of ambipolar diffusion as this provides 
a natural explanation for the differential behavior of velocity anisotropy between the 
cloud envelope and the high column density filaments.
 
\section{Conclusions} 
We have analyzed  \co\ and \coa\ J=1-0 emissions within equally partitioned areas of the Taurus molecular cloud
for velocity anisotropy.   Significant anisotropy aligned with the local magnetic field direction is limited to 
the envelope of the cloud as traced by low surface brightness \co\ emission.  The \coa\ emission, which 
effectively probes the dense filaments, exhibits little evidence for such anisotropy.  This 
differential behavior between the low column density envelope and the dense interior results from a 
transition from sub-Alfv\'{e}nic turbulence to super--Alfv\'{e}nic motions.  
Such a transition is realized if the magnetic field is not amplified with increasing density as expected from ambipolar diffusion or forced loading of field lines 
by magnetically aligned cloud-cloud collisions.  

\section*{Acknowledgments}
MH is supported by grants AST-0838222 and AST-1009049 from the National Science Foundation.

\clearpage
\clearpage
\clearpage
\clearpage
\clearpage

\end{document}